\documentclass[12pt,preprint]{aastex} 
\usepackage{emulateapj5}
\usepackage{epsfig}
\usepackage{xspace}
\usepackage{natbib}
\usepackage{amsmath}

\newcommand{\sS}[1]{\mbox{$\rm{}^{#1}$}}
\newcommand{\Ss}[1]{\mbox{$\rm{}_{#1}$}}

\renewcommand{\arcdeg}{\mbox{$^\circ$}\xspace}

\newcommand{\PLI}{\mbox{$\Gamma$}\xspace}
\newcommand{\nH}{\mbox{$N\Ss{H}$}\xspace}
\newcommand{\dqt}[1]{\mbox{$Q_{#1}$}\xspace}
\newcommand{\qt}[1]{\mbox{$E_{#1\%}$}\xspace}

\begin{document}

\title{New spectral classification technique 
for X-ray sources: quantile analysis}

\slugcomment{Received: 2 Feb 2004, Accepted: \_\_\_\_\_\_\_\_\_\_\_\_}

\author{
Jaesub Hong\altaffilmark{1}, 
Eric M. Schlegel\altaffilmark{1}, and Jonathan E. Grindlay\altaffilmark{1}
}
\email{jaesub@head.cfa.harvard.edu}
\altaffiltext{1}{Harvard-Smithsonian Center for Astrophysics, 60 Garden St.,
Cambridge, MA 02138}

\begin{abstract}

We present a new technique called {\it quantile analysis} to classify spectral
properties of X-ray sources with limited statistics. The quantile
analysis is superior to the conventional approaches such as X-ray
hardness ratio or X-ray color analysis to study relatively faint sources
or to investigate a certain phase or state of a source in detail, where
poor statistics does not allow spectral fitting using a model.  Instead
of working with predetermined energy bands, we determine the energy
values that divide the detected photons into predetermined fractions of
the total counts such as median (50\%), tercile (33\% \& 67\%), and
quartile (25\% \& 75\%).  We use these quantiles as an indicator of the
X-ray hardness or color of the source.  We show that the median is an
improved substitute for the conventional X-ray hardness ratio. The
median and other quantiles form a phase space, similar to the
conventional X-ray color-color diagrams.  The quantile-based phase space
is more evenly sensitive over various spectral shapes than the
conventional color-color diagrams, and it is naturally arranged to
properly represent the statistical similarity of various spectral
shapes.   We demonstrate the new technique in the 0.3-8 keV energy range
using Chandra ACIS-S detector response function and a typical aperture
photometry involving background subtraction.  The technique can be
applied in any energy band, provided the energy distribution of photons
can be obtained.

\end{abstract}


\section{Introduction}

A major obstacle for understanding the nature of the faintest X-ray
sources is poor statistics, which prevents the usual practice of
spectral analysis such as fitting with known models.  Even for
apparently bright sources, as we investigate the sources in detail, we
frequently need to divide source photons in various phases or states of
the sources and the success of such an analysis is often limited by
statistics.  The common practice to extract spectral properties of X-ray
sources with poor statistics is to calculate X-ray hardness or color of
the sources \citep{Schulz89,Kim92,Netzer94,Prestwich03}.  In this
conventional method, the full-energy range is divided into two or three
sub-bands and the detected source photons are counted separately in each
band.  The ratio of these counts is defined as X-ray hardness or color,
to serve as an indicator of the spectral properties of the source.

In principle, one can constrain a meaningful X-ray hardness or color to
a source if at least one source photon is detected in each of at least
two bands. However, the equivalent requirement for total counts in the
full-energy band can be rather demanding and it is strongly spectral
dependent.  Fig.~1 shows an example of a X-ray color-color diagram
\citep{Kim03}.  In the figure, we divide the full-energy range
(0.3-8.0~keV) into three sub-energy bands; 0.3-0.9 (S), 0.9-2.5 (M), and
2.5-8.0~keV (H).  The ratio of the net source counts in S to M band is
defined as the soft X-ray color ($x$-axis in the figure), and the ratio
of the net counts in M to H band as the hard X-ray color ($y$-axis).
The energy range in this example is the region where the Chandra ACIS-S
detectors are sensitive, but the technique described in this paper is
not bounded to any particular energy range, provided the energy
distribution of photons can be obtained.

The grid pattern in the figure represents the true location for the
sources with the power-law spectra governed by two parameters: power-law
index (\PLI) and absorbing column depths (\nH) along the line of the
sight. The grid pattern is drawn for an ideal detector response, which
is constant over the full-energy band.  The grid pattern appears to be
properly spaced to the changes of the spectral parameters and such an
arrangement suggests that this color-color diagram may be an ideal
way to classify sources with poor statistics.  

However, the appearance can be deceiving, which is hinted by the uneven
sizes of error bars in the figure.  The error bar near each grid node
represents the central 68\% of simulation results from the spectral
shape at the grid node.  Each simulation has 1000 source counts in the
full-energy range with no background counts, and thus the size of the
symbol represents the size of typical error bars for 1000 count sources
in the diagram.

In the figure, there are two kinds of error bars for some spectral
shapes (e.g. \PLI= 0 and \nH= $10^{22}$ cm\sS{-2}); the thick error bars
represent the central 68\% distribution of the simulation results that
have ``proper'' color values (defined here as at least one photon in
each of all three bands), and the thin error bars show the 68\%
distribution of all the simulation trials (10,000) for each spectral
model, i.e. a distribution of the central 6827 trials. The two error
bars should be identical if all the trials produce a proper soft and
hard color.

The right panel in Fig.~1 shows the minimum counts in the full-energy
band to have at least one count in each of three bands.  The figure
indicates that the required minimum counts are dramatically different
among the models in the grid.  For the power-law spectra with \PLI= 0
and \nH $\sim 10^{22}$ cm\sS{-2}, more than 500 counts are required in
order to have reasonable color values, while in the case of \PLI= 2  and
\nH $\sim 10^{20}$ cm\sS{-2}, a few counts are sufficient.  Because of
statistical fluctuations, for some models, even 1000 counts in the
full-energy range do not guarantee positive net counts in all three
bands, which explains why some trials fail to produce proper colors.

The spectral-model dependence of error bars in this kind of color-color
diagram is inevitable and the dependence is determined by the choice of
the sub-energy bands for a full-energy range. In this example, the sub
bands are chosen so that the diagram is mostly sensitive near \PLI= 2
and \nH $< 10^{22}$ cm\sS{-2}. In fact, it is not possible to select sub
bands to have uniform sensitivity over different spectral shapes.

Fig.~1 clearly indicates that the conventional color-color diagram
(CCCD) is heavily biased, which is very unfortunate when trying to
extract unknown spectral properties from the sources. The heavy
requirement on the total counts for certain spectral shapes defeats the
purpose of the diagram since the large total counts may allow 
nominal spectral analysis such as spectral fitting with known models.
One might have to repeat the analysis with different choices of
sub-bands to explore all the interesting possibilities of spectral
shapes.

\begin{figure*} \begin{center} 
\epsscale{1.0}
\plotone{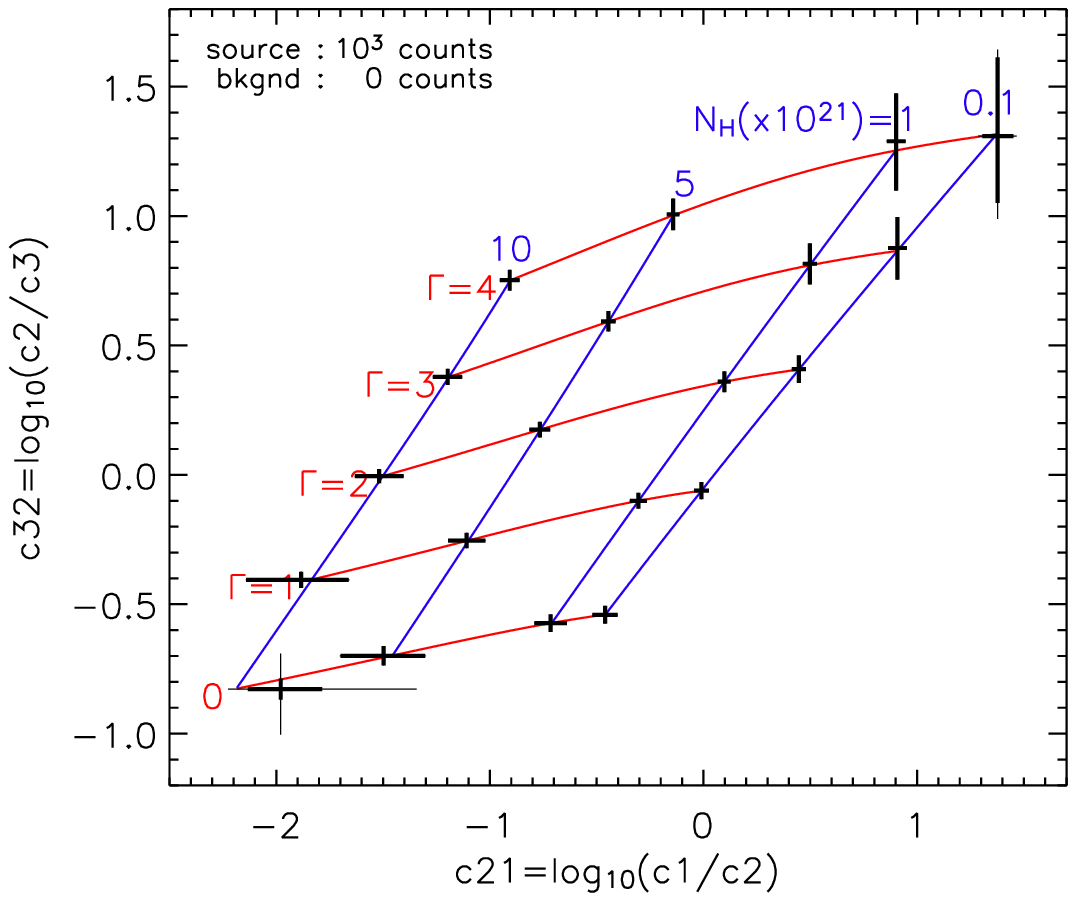}
\plotone{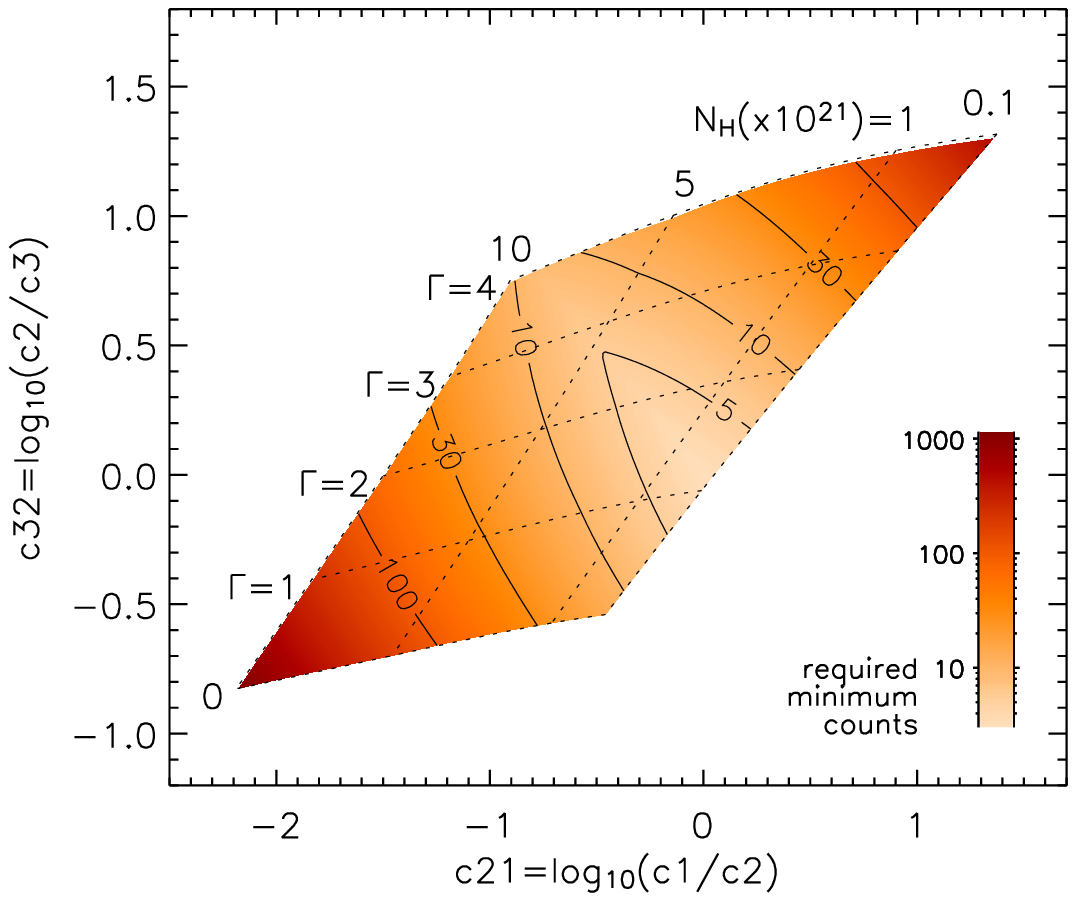}
\end{center} \caption{Conventional color-color diagram (CCCD): two colors are
defined by the net counts (c1, c2, and c3) in three energy bands
(0.3-0.9, 0.9-2.5, and 2.5-8.0 keV).  The left panel shows simulation
results of 20 spectral shapes for an ideal detector with a uniform
response function.  The error bar represents the central 68\% of
simulation results for each of 20 power-law spectra governed by two parameters
-- power-law index (\PLI) and column depth (\nH) along the line of the
sight.  The thick error bars are for the 68\% of the cases
with proper color values (at least one count in each of the three
bands) and the thin error bars for considering 10,000
trials (see text).  Each simulation run contains 1000 source counts in
the full 0.3-8.0 keV energy range.   In this example, the two error bars
are identical except for the case of \PLI=0 \& \nH=10\sS{22} cm\sS{-2}
and \PLI=4 \& \nH=10\sS{20} cm\sS{-2}, where many simulation runs result
in null count in at least one of sub-bands.  The grid pattern represents
the true location of these spectra in the diagram.  The right panel
shows the required minimum counts in the full-energy band in order to
have at least one count in each of three bands. The contour lines
quantize the gray (color) scale for easier interpretation.
See electronic ApJ for color version of the figure as well as other
figures.} \label{fig:cc_pl}
\end{figure*}

\section{Quantiles}

In order to overcome the selection effects originating from the
predetermined sub-energy bands, we propose to use the energy value to
divide photons into predetermined fractions. We choose fractions to take
full advantage of the given statistics, such as 50\% (median), 33\% \&
67\% (tercile), and 25\% \& 75\% (quartile), although any quantile may
be used. We use the corresponding energy values - quantiles - as an
indicator of the X-ray hardness or color of the source.  In particular,
we will show below that the median, \dqt{50}, is an improved substitute
for the conventional X-ray hardness.

Let \qt{x} be the energy below which the net
counts is $x\%$ of the total counts and we define quantile \dqt{x}
\begin{eqnarray*}
	\dqt{x} = \frac{\qt{x} - E\Ss{lo}}{E\Ss{up}-E\Ss{lo}},
\end{eqnarray*}
where $E\Ss{lo}$ and $E\Ss{up}$ is the lower and upper boundary of the
full-energy band respectively (0.3 and 8.0 keV in this example).  
The algorithm for estimation of
quantiles and their errors relevant to X-ray astronomy is given in the
appendix\footnote{The quantile analysis software ({\tt IDL} and {\tt
perl} versions) is available at {\tt
http://hea-www.harvard.edu/ChaMPlane/quantile}.}.  
Unlike the conventional X-ray hardness or colors, for
calculating quantiles,  there is no spectral dependence of required
minimum counts.  The required minimum only depends on types of
quantiles: two counts for median and terciles, and three for quartiles.

Fig.~2 shows an example of quantile-based color-color diagrams (QCCDs)
using the median $m (\equiv\dqt{50})$ for the $x$-axis and the ratio of
two quartiles \dqt{25}/\dqt{75} for the $y$-axis (see \S 5 for the
motivation of the choice of the axes).  The grid pattern in Fig.~2 is
drawn for the same spectral parameters as in Fig.~1 with five additional
cases for \nH= $5 \times 10^{22}$ cm\sS{-2}.  Note that the approximate
axes of \PLI and \nH are rotated $\sim$ 90$\deg$ from those in Fig.~1.
The error bars are drawn for the central 68\% of the same 10,000
simulation results of the spectral shape at each grid node, and each
simulation run contains 1000 source counts with no background counts in
the full-energy band.  The relatively similar size of the error bars for
1000 count sources in the figure indicates that there is no spectral
dependent selection effect.  Note that there is no need of distinction
for thin and thick error bars in this figure because all the trials
produce proper quantiles.

The grid pattern of power-law spectra in Fig.~2 appears to be less
intuitively arranged than the one in Fig.~1.  However, we believe that
the proximity of any two spectra in the quantile-based diagram
accurately exhibits the similarity of the two spectral shapes (as folded
through the detector response, which is constant in this example).  For
example, in the case for \PLI= 0 in Fig.~2, the separation in phase space
by various \nH values is much smaller than in the case for \PLI$\ge$ 1.
Note that the column depth (\nH) in the considered range can change
mainly soft X-rays ($\lesssim$ 2 keV) and the spectrum for \PLI= 0 is
less dominated by soft X-rays than that for \PLI$\ge$ 1.  In other words,
the overall effects of \nH on the spectral shape would be much smaller
in the case for \PLI= 0 than in the case for \PLI$\ge$ 1.  Therefore, the
grid spacing in Fig.~2 indeed reveals the appropriate statistical
power needed to discern these spectral shapes, despite the degeneracy arising
from utilizing only three variables (quantiles) extracted from a spectrum.

\begin{figure*} \begin{center} 
\epsscale{1.0}
\plotone{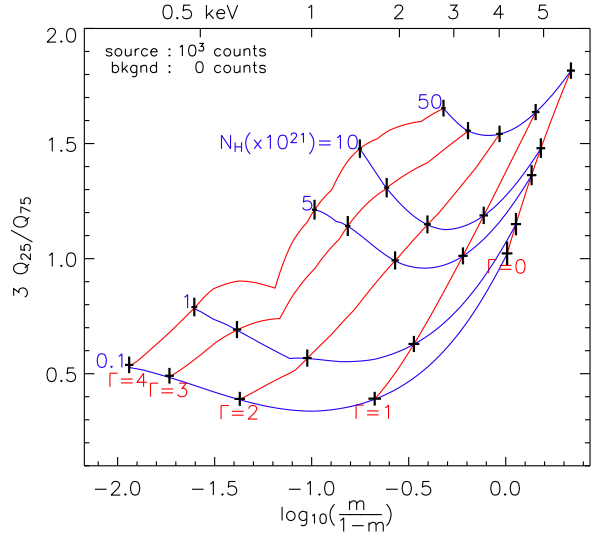} \end{center} \caption{Same as Fig.~1 but quantile-based
color-color diagram (QCCD) based on the median \dqt{50} and the ratio of two
quartiles \dqt{25}/\dqt{75}. See \S 5 for the origin of 
the logarithmic form for the $x$-axis. 
The energy scale across the top shows the median energy values (\qt{50}).
The simulation results are
given from the same simulation sets in Fig.~1 (1000 source counts). 
Note that there are five more spectral shapes with \nH= $5\times 10^{22} $
cm\sS{-2} than in Fig.~1.  The grid pattern represents the true
location of these spectra in the diagram for an ideal detector. Note
the relative uniformity of error bars, compared to Fig.~1. } \label{fig:c2_pl} \end{figure*}

\section{Color-color diagram example}

Now let us consider more realistic examples. We introduce the Chandra
ACIS-S response function\footnote{See Chandra Proposers' Observatory
Guide at {\tt http://cxc.harvard.edu/}.}; we use the pre-flight
calibration data for energy resolution\footnote{$\sim$ 100 -- 250 eV
FWHM. See {\tt http://cxc.harvard.edu/cal/Acis/}.} and the energy
dependent effective area from PIMMS version 2.3\footnote{Portable,
Interactive Multi-Mission Simulator. See {\tt
http://heasarc.gsfc.nasa.gov/docs/software/tools/pimms.html}.}.

Fig.~3 shows the simulation results of the spectra with 100 and 50
source counts with no background counts. The error bars again indicate
the interval that encloses the central 68\% of the simulation results for
each model. The results are shown for both the conventional and new
color-color diagrams in the figure.  The grid patterns in the figures
are for the same power-law models and they look different from the
previous examples because of the Chandra ACIS-S response function.

The analysis of faint sources is often limited by background
fluctuations. In contrast to Fig.~3, we allow for large background
counts in the source region (e.g. a point source in a bright diffuse
emission region).  The top two panels in Fig.~4 shows the simulation
results of the spectra with 100 source counts and 50 background counts
in the source region, and the bottom two panel shows the case of 50
source counts and 25 background counts.   For background subtraction, we
set the background region to be five times larger than the source
region.  The background region contains 250 (top panels) and 125
background (bottom panels) counts respectively.  Note these are
relatively high backgrounds for a typical Chandra source and thus
illustrate the worst case of a point source superimposed in background
diffuse emission.  The background photons are sampled from a power-law
spectrum with \PLI= 0 and \nH= 0, which is folded through the Chandra
ACIS-S response function.

In the case of the CCCD in Figs.~3 and 4, one
can notice that some of the error bars lie away from their true location
(grid node). The severe requirement for the total counts ($>100$) for
some of the spectral models results in only a lower or upper limit for
the color in many simulation runs. Cases with proper colors for these
models are greatly influenced by statistical fluctuations because of low
source counts in one or two sub-bands, and thus the estimated colors
fail to produce the true value.  It is evident that this conventional
diagram is more sensitive towards \PLI$\sim$ 2 and \nH $\lesssim
10^{22}$ cm\sS{-2}.

In the case of the new quantile-based diagram, the error bars stay at
the correct location and the size of the error bars are relatively
uniform across the model phase space, indicating that the phase space is
properly arranged. The quantile-based diagram shows more consistent
results regardless of background.

\begin{figure*} \begin{center} 
\epsscale{1.0}
\plotone{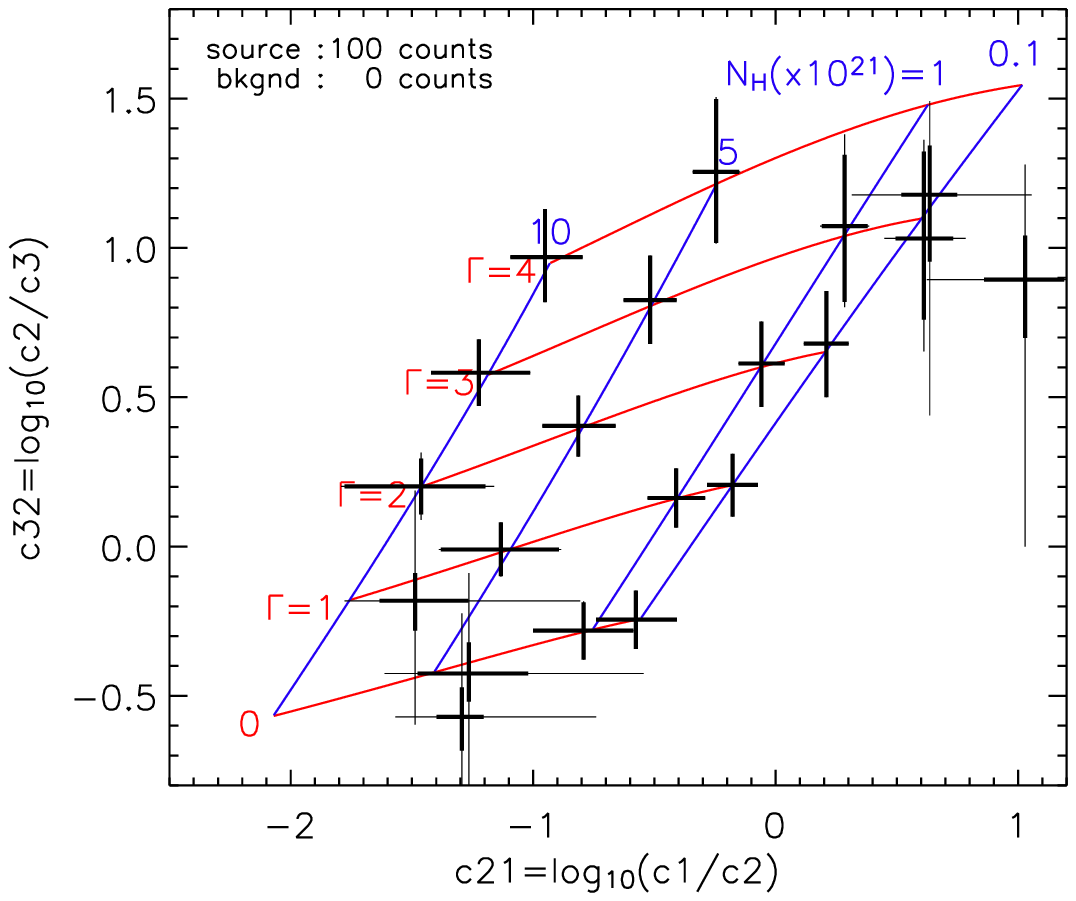}
\plotone{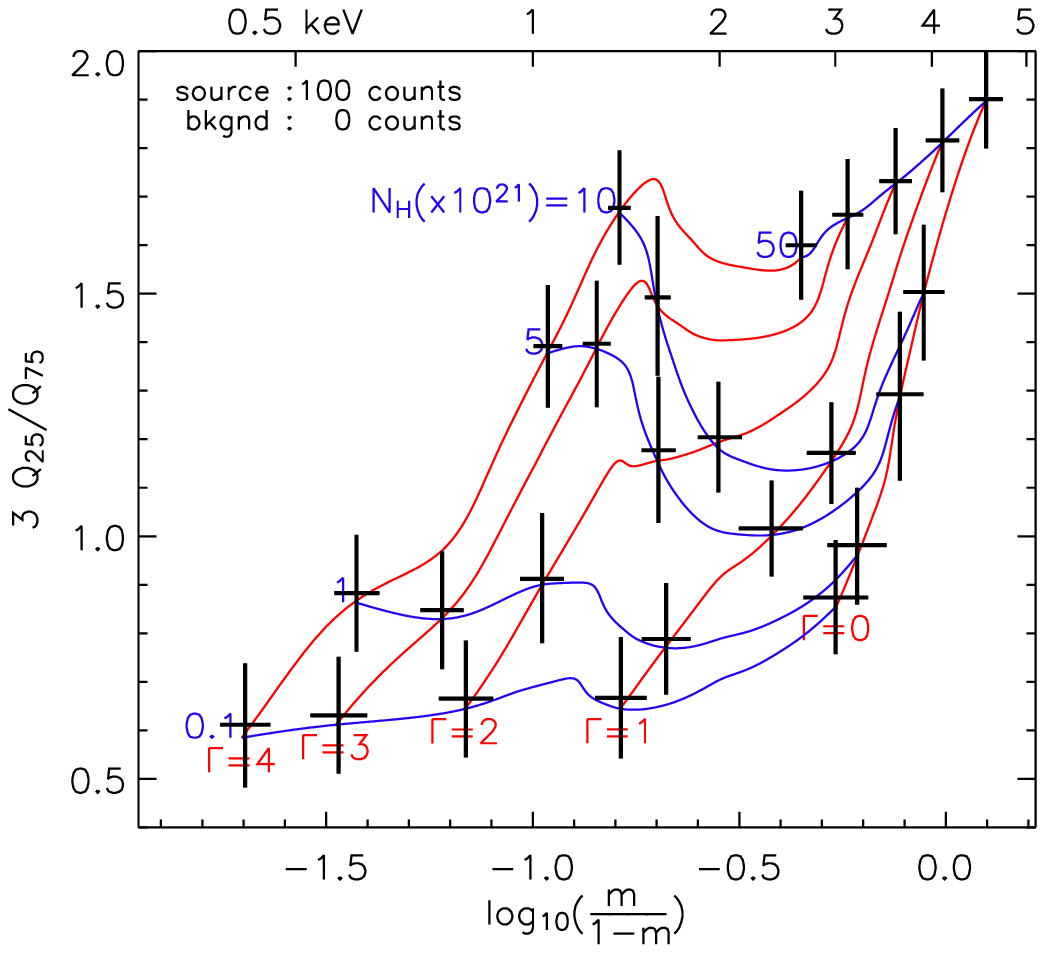}
\plotone{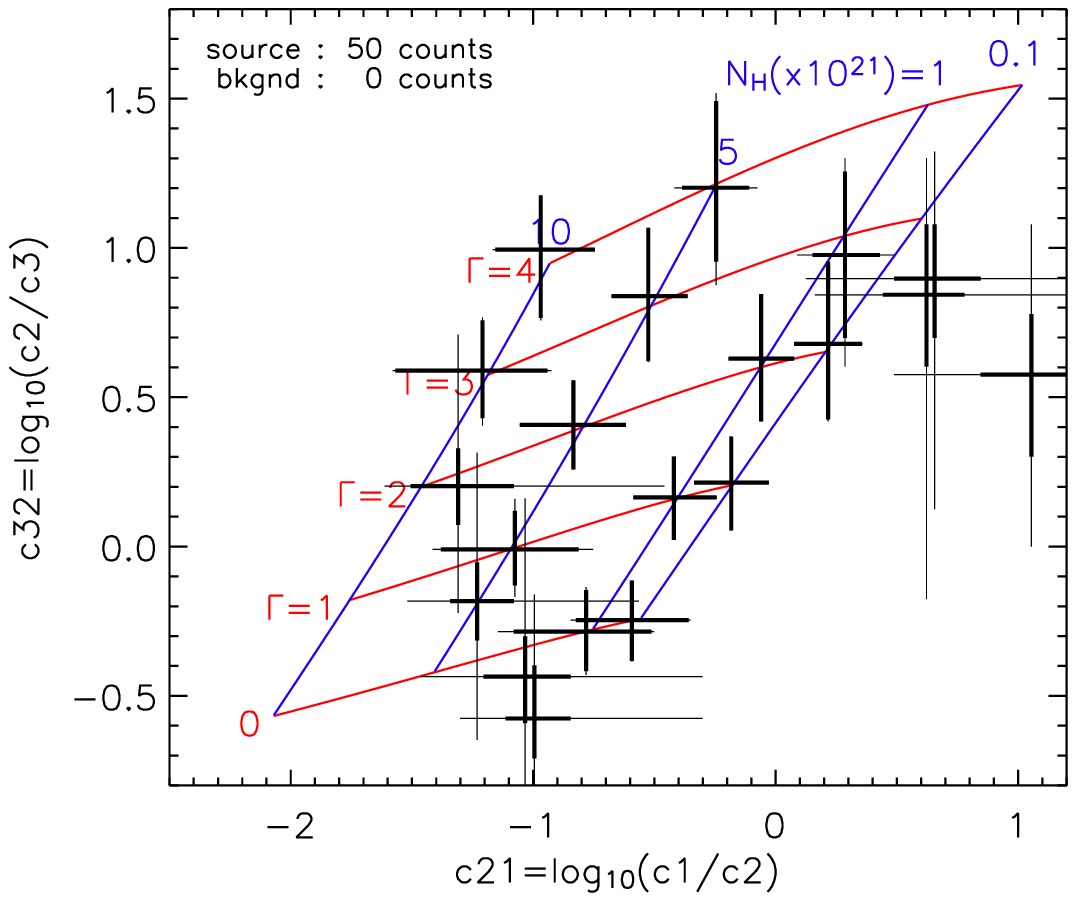}
\plotone{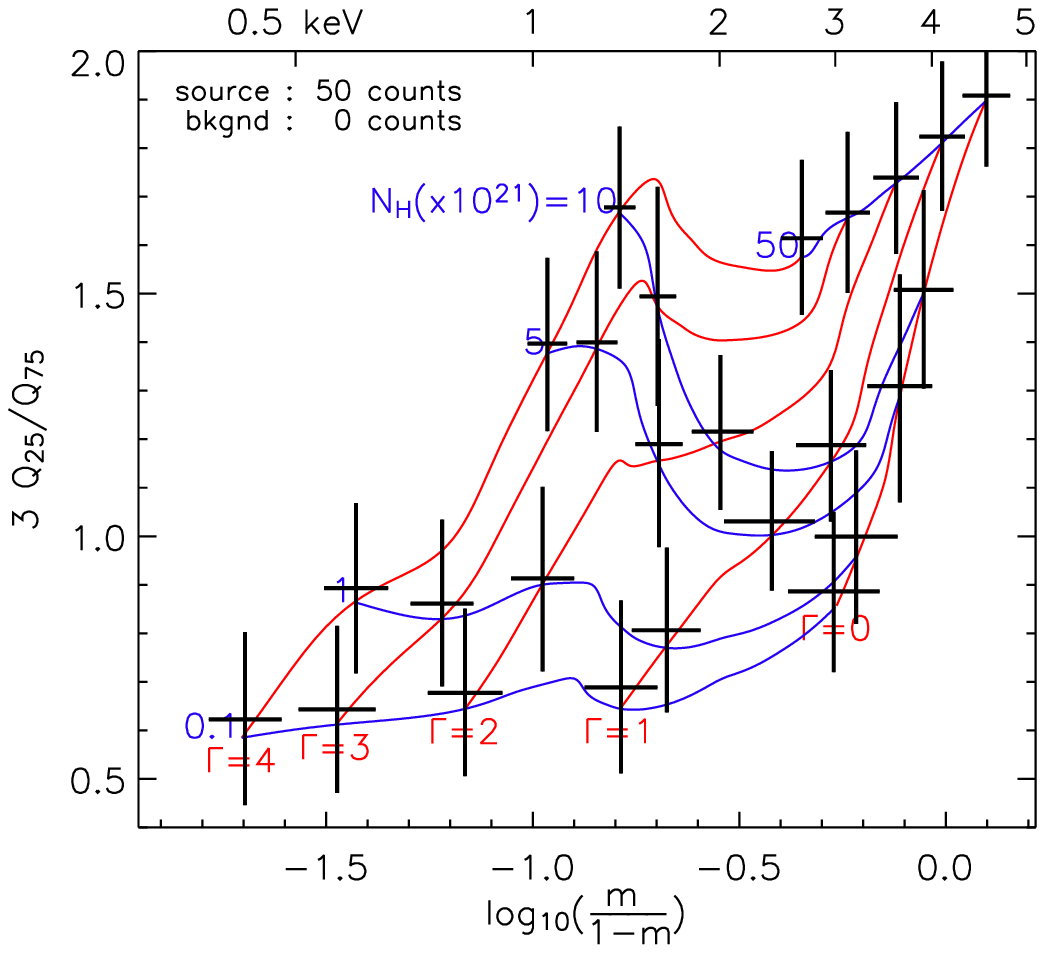}
\end{center} 
\caption{Realistic example of the color-color diagram with no
background: 100 (top panels) and 50 (bottom panels) source
counts for each spectral model folded through the Chandra ACIS-S
response function.  The left two panels show the CCCD
and the right two panels show the QCCD.
The grid pattern of both cases are different from the previous example
because of the ACIS-S response function. 
In the case of the CCCD, the average of the
simulation results are often severely different from the true value
(grid node) due to the selection effect inherent to the band choice. 
The selection effect also causes the large difference between thick
and thin error bars, which indicates that only a fraction of simulations
produce proper colors. The fraction can be even smaller than 
68\% of the total 10,000 runs, in which case the thin error bar represents 
the full range of the simulation outcomes with proper colors.  The
quantile-based diagram provides the correct results regardless of
spectral shapes. }
\label{fig:pl_aciss_nobkg}
\end{figure*}

\begin{figure*} \begin{center}
\epsscale{1.0}
\plotone{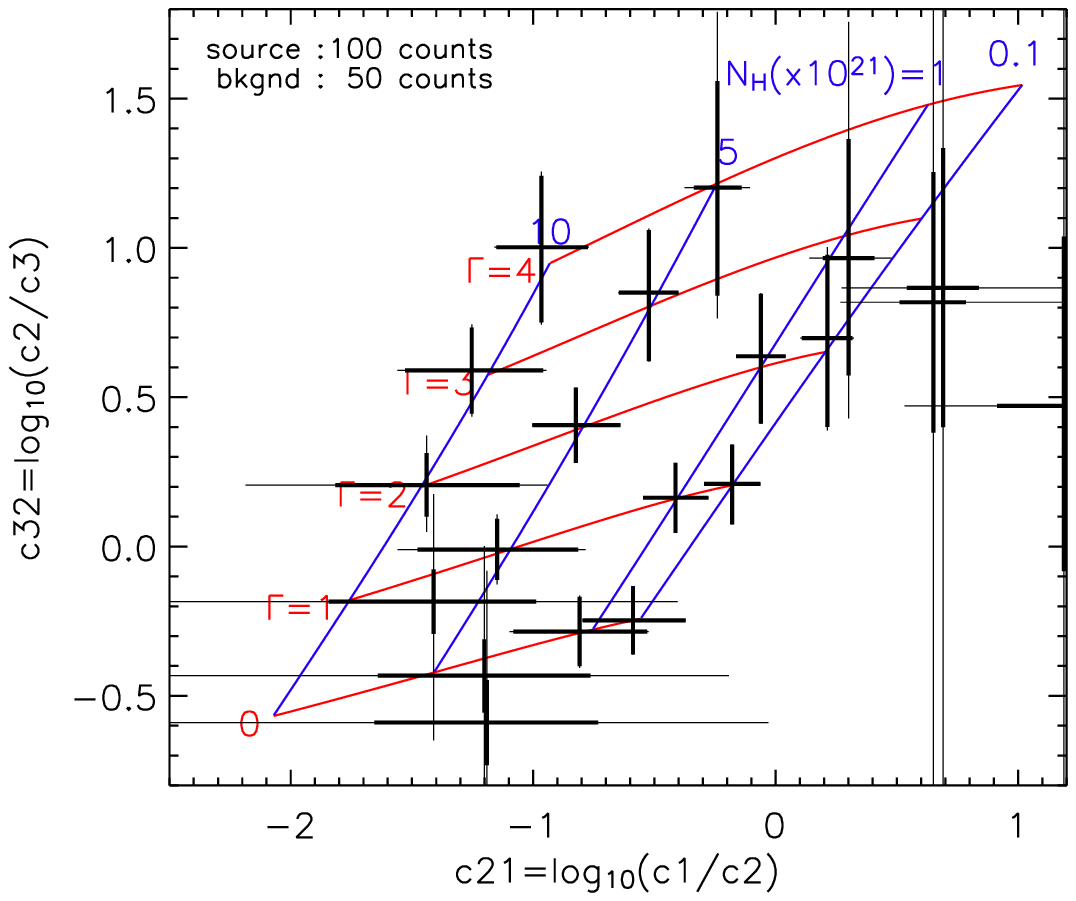}
\plotone{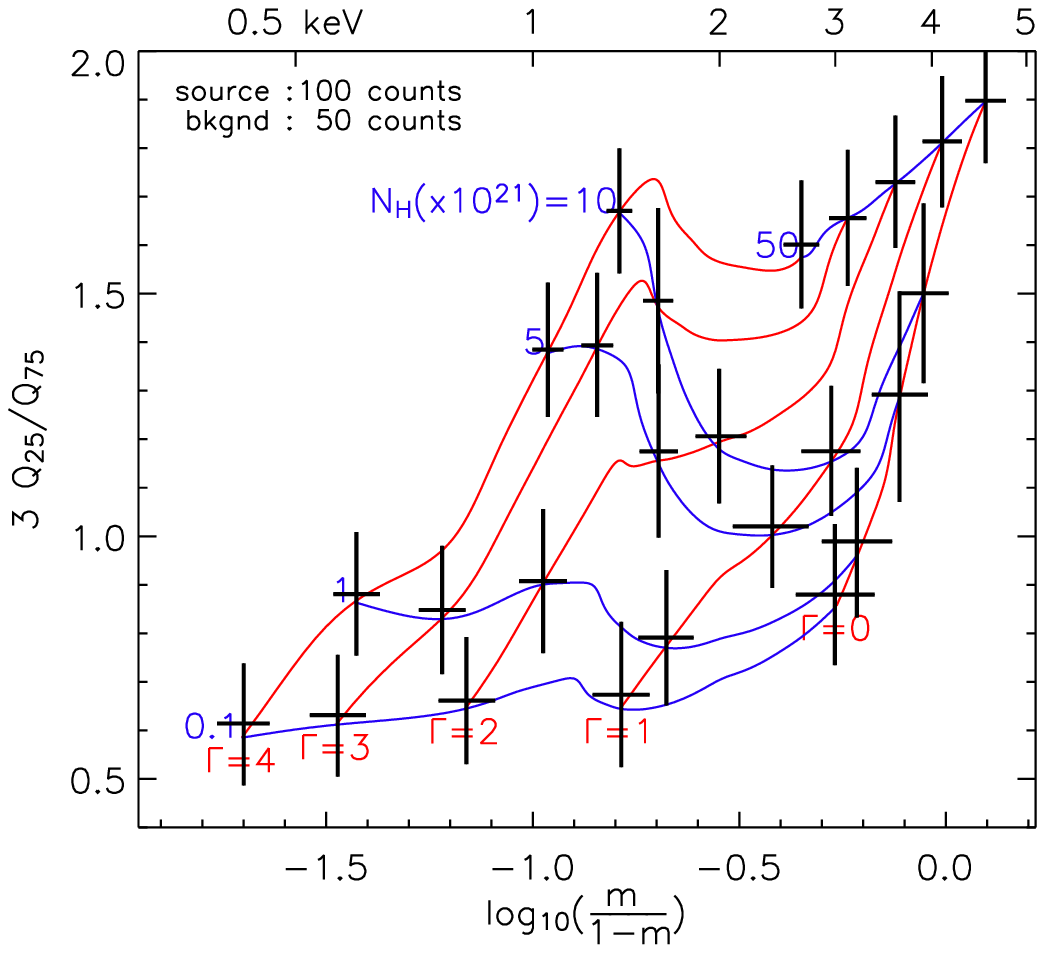}
\plotone{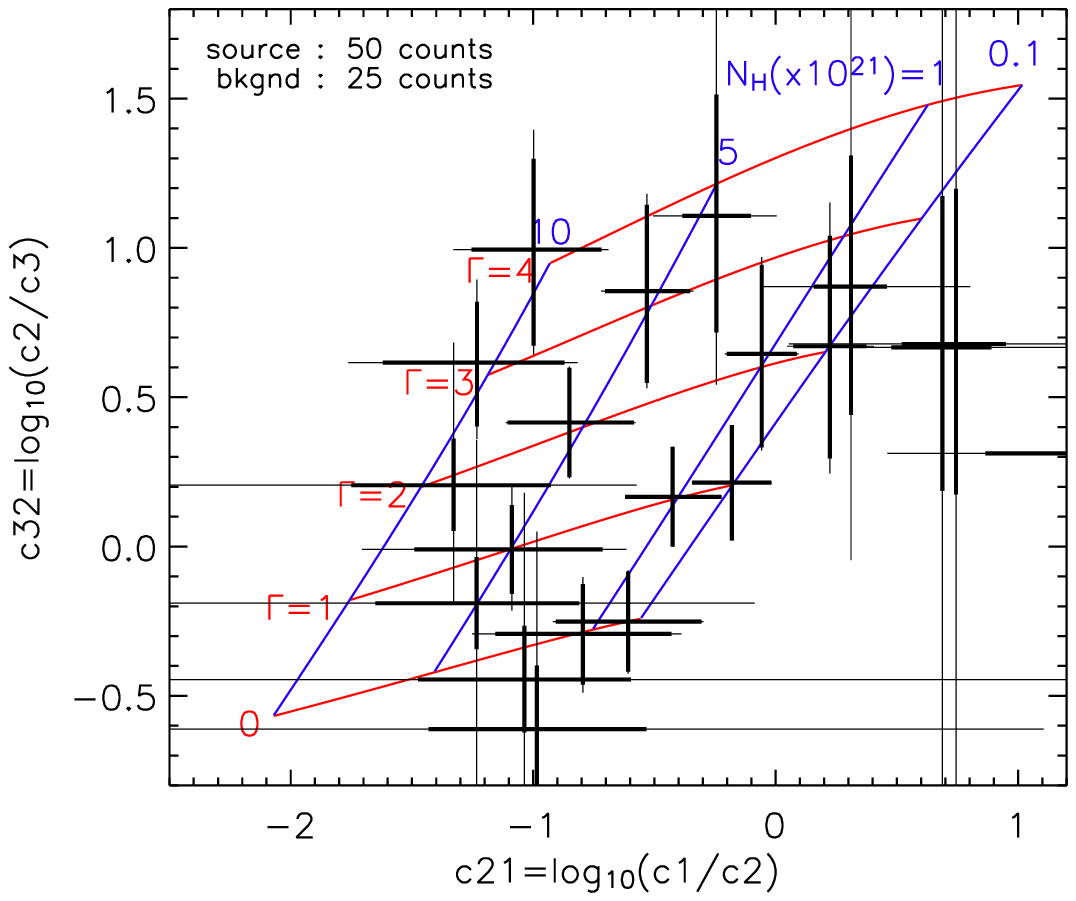}
\plotone{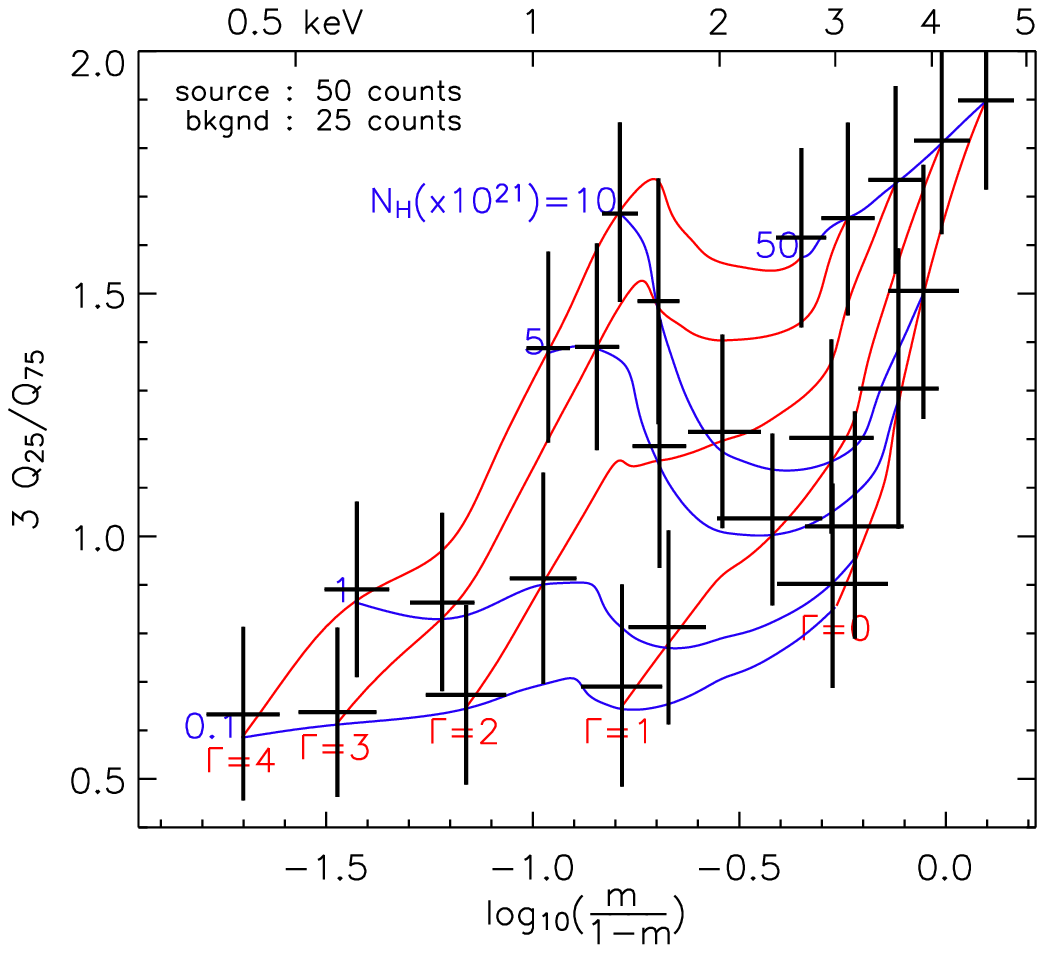}
\end{center} \caption{Realistic
example of the color-color diagram with relatively high background: 100 count source
with 50 background counts (top panels) and 50 count source with 25
background counts (bottom panels). The left two panels show the
CCCD, and the right two panels show the
QCCD. The background follows a flat spectrum, and
both source and background photons are folded through the Chandra
ACIS-S response function. The diagrams are generated after the
background subtraction (see the appendix).  
Similar to the previous
results, the new technique performs well throughout the phase space. 
In particular, note the relatively uniform error bars across the entire
grid as well as the relative lack of points outside the grid.
}
\label{fig:pl_aciss_bkg}
\end{figure*}

\section{Hardness ratio example}

When we explore the spectral properties of sources with an even smaller
number of detected photons, the only available tool is the use of a single
hardness ratio, which requires only two sub-energy bands.  
Here we use net counts $S$ in a
0.3-2.0 keV soft band and net counts $H$ in a 2.0-8.0 keV hard band. Two
popular definitions for hardness ratio exist in the literature: HR =
$H/S$ and HR=$(H-S)/(H+S)$. We compare the median with the conventional
hardness ratio using the first definition.

The left panel in Fig.~5 shows the performance of the conventional
hardness ratio as a function of the total counts folded through the
ACIS-S response. The plot contains three spectral shapes with \PLI= 2.
The shaded regions represent the central 68\% of the
simulation results for a given total net counts (except for the case of
$H$ or $S$ $\le$ 0). The simulation is done for the case without
background (top panels) and for the case that the source region contains
an equal number of source and background counts (bottom panels), where
we perform a background subtraction identical to the one described in
the previous section. The background follows the same spectrum as in the
previous examples.

Similar to the CCCD, the required minimum of
the total counts for the conventional hardness ratio depends on spectral
shape. The right panels in Fig.~5 show the fraction of the simulations
that have at least one count in both the soft and hard bands (otherwise
HR = 0 or $\infty$). The plot indicates that one cannot assign a proper
hardness ratio value in a substantial fraction of cases when the total
net counts are less than 100.  In the case of $\nH= 10^{20}$ cm\sS{-2},
many simulation runs result in $H=0$ (too soft), and for $\nH= 5 \times
10^{22}$ cm\sS{-2}, $S=0$ (too hard). A set of predetermined bands keeps
the hardness ratio meaningful only within a certain range of spectral
shapes.  In order to compensate for such a limitation, one needs to
repeat the analysis with different choices of bands, which in turn will
have a different limitation.

Even for the cases with a proper hardness ratio value (the left panel in
Fig.~5), the hardness ratio distribution of two spectral types (\nH=
$10^{22}$ and $5\times 10^{21}$ cm\sS{-2}) overlaps
substantially when the total net counts are less than 40 in the case of
high background. Below 10 counts, it is difficult to relate the
distributions to their true value.

\begin{figure*} \begin{center} 
\epsscale{1.0}
\plotone{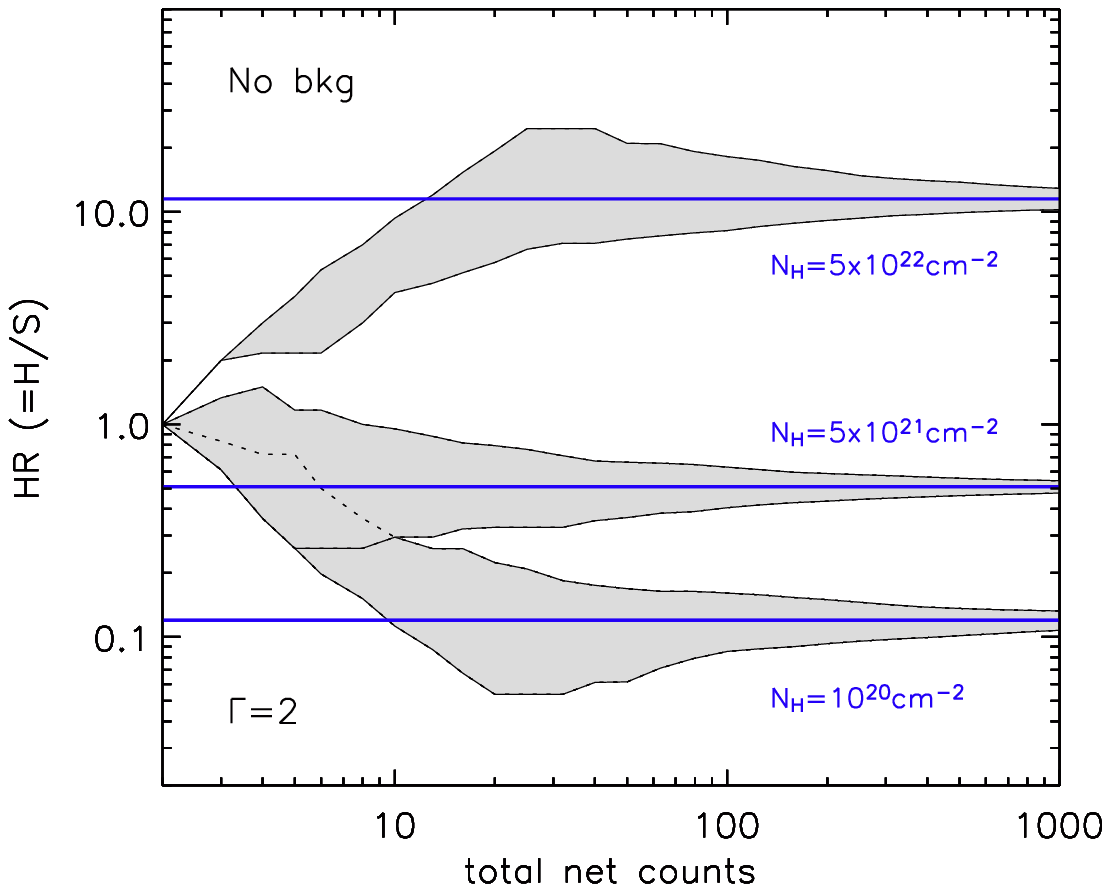}
\plotone{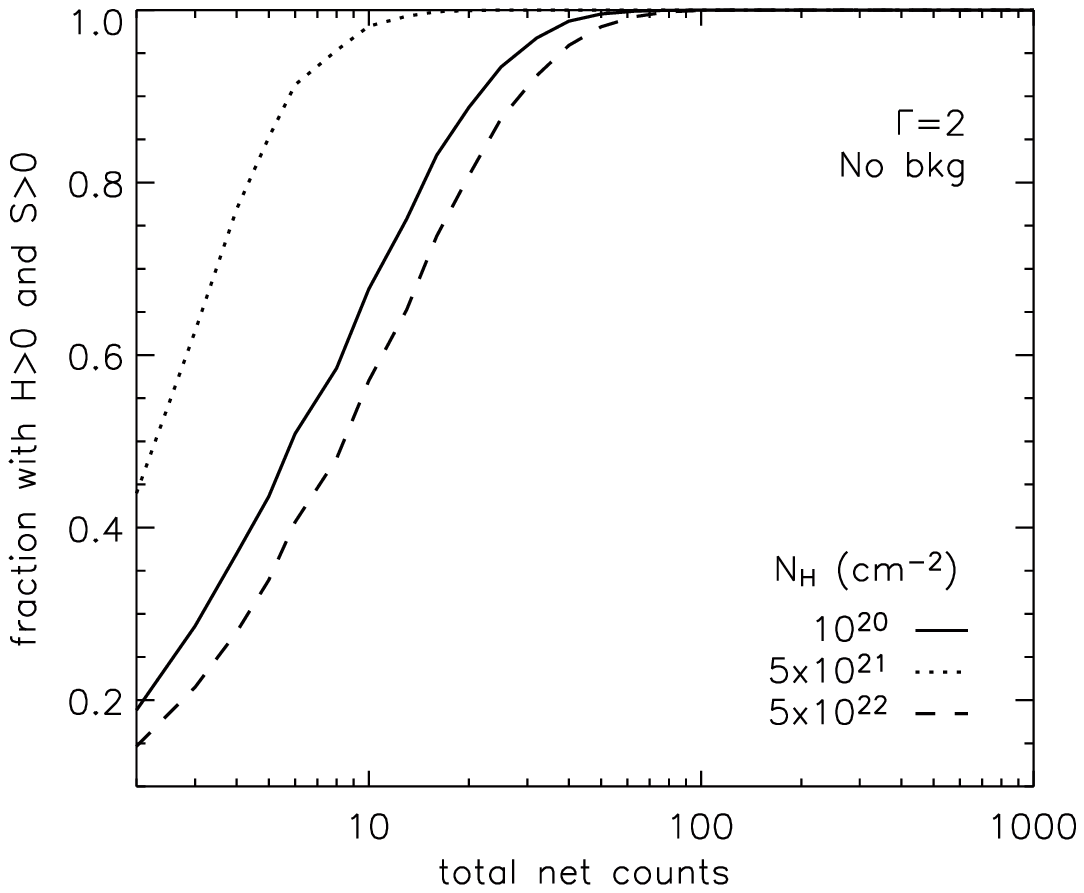}
\plotone{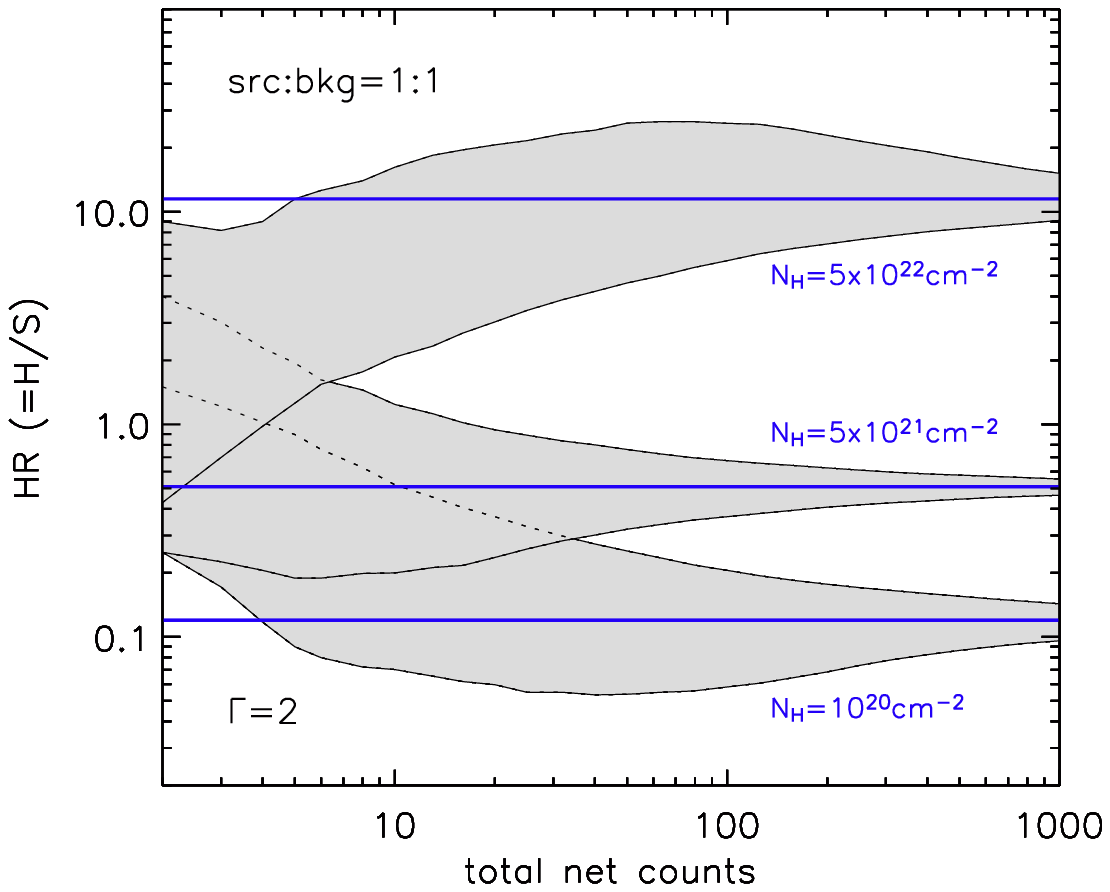}
\plotone{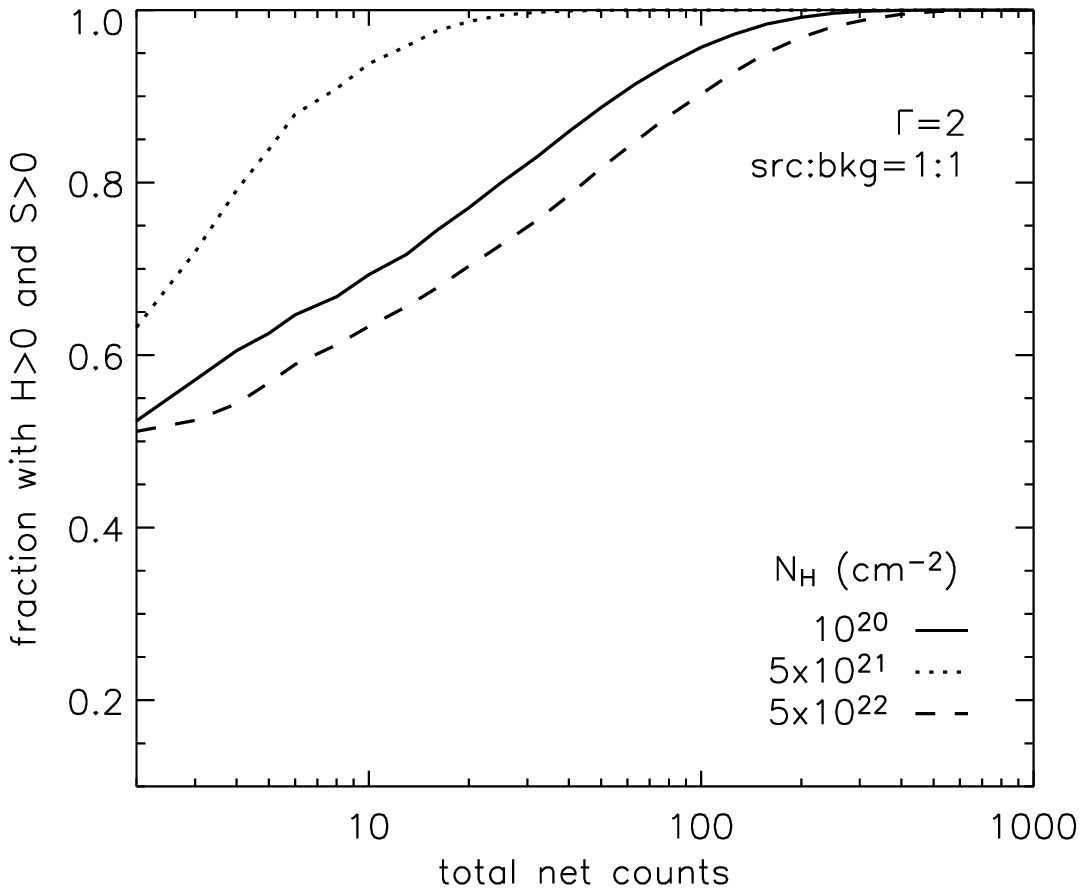}
\end{center}
\caption{Conventional hardness ratio calculation. The soft
band ($S$) is 0.3-2.0 keV and hard band ($H$) 2.0-8.0 keV, and the Chandra ACIS-S
response function is used.
The left two panels show a conventional hardness distribution as a
function of total net counts and the right two panels show the fraction
of the simulations that have $H>0$ and $S>0$ for a given total net
count.  The top two panels are for the case of no background, and in the
bottom two panels, for a given net count, the source region contains an
equal number of source and background counts, and we perform a background
subtraction identical to the one described in the previous examples.
Three spectral shapes (\PLI= 2) are simulated, and the solid horizontal
lines in the left panels represent the true hardness ratio of each spectrum.  
The shaded
region represents the central 68\% of simulated results (only
for $H>0$ and $S>0$) at a given total net count.  
A large fraction of trials do not produce a proper hardness ratio when
the total counts are less than 100.  Below 10 counts, the hardness
distribution hardly bears any relation to its true value.  Blending
of the central 68\% distributions starts at $\sim$ 40 counts for \nH= $10^{20}$ and
$5\times10^{21}$ cm\sS{-2} for the case with background.}
\label{fig:conventionalHR} \end{figure*}

Fig.~6 shows the performance of the new hardness defined by the median.
First, there is no loss of events. Second, three spectra are well
separated down to less than 10 counts regardless of background.  Even at
two or three net counts, each distribution is distinct and well related
to its true value.

\begin{figure*} \begin{center} 
\epsscale{1.0}
\plotone{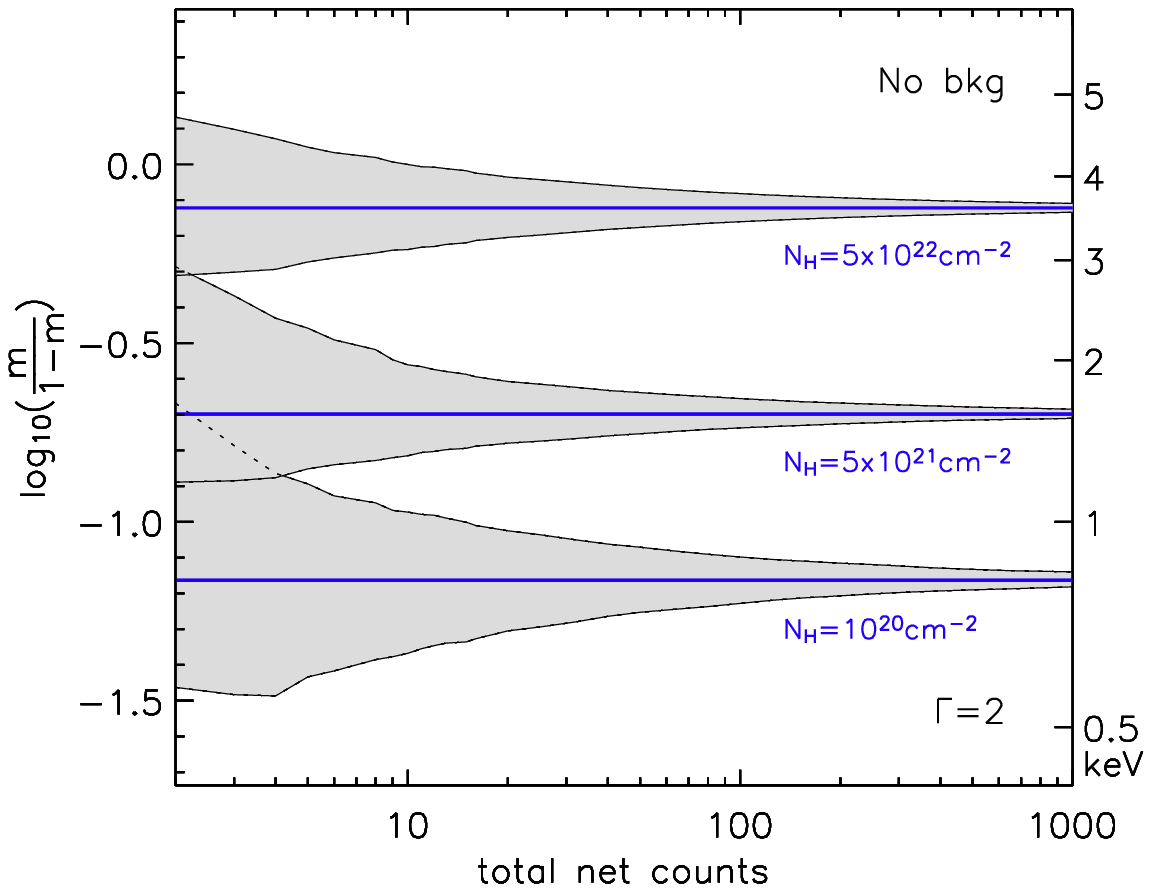}
\plotone{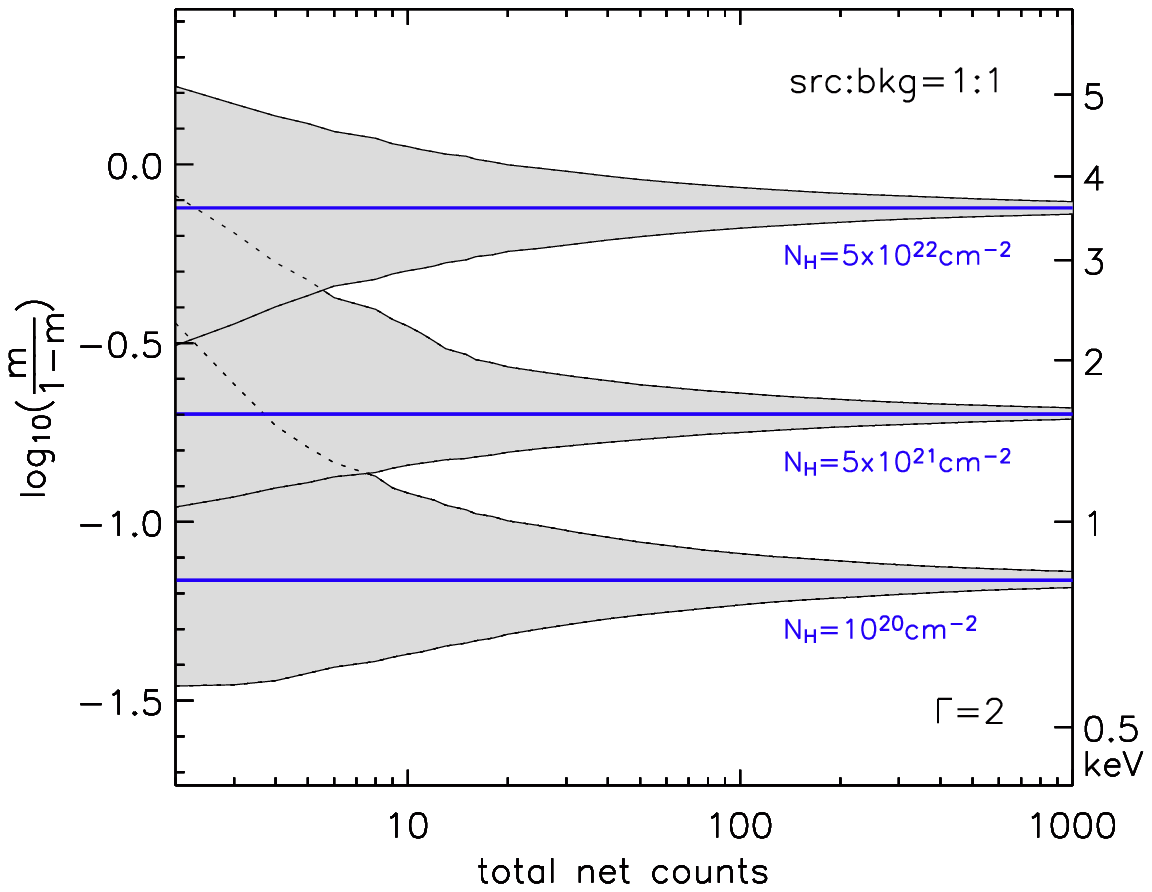}
\end{center} \caption{The median
distribution for three power-law spectra with 
$\Gamma = 2$. Compare this with Fig.~5. Regardless of background,  
the three distributions are well separated down to less than 10 counts,
bearing a tight relation with their true value.  In addition, there is
no loss of events in this diagram.}
\label{fig:newHR} \end{figure*}

\section{Discussion}

The newly defined hardness and color-color diagrams using the median and
other quantiles are far superior to the conventional ones. One can
choose specific quantiles and their combinations relevant to one's
needs.  We believe the new phase space by the median and quartiles is
very useful in general, as summarized in Fig.~7.

For a given spectrum, various quantiles are not independent variables,
unlike the counts in different energy bands.  However, the ratio of two
quartiles is mostly independent from the median, which makes them good
candidate parameters for color-color diagrams. In essence, the median \dqt{50}
shows how dense the first (or second) half of the spectrum is and the
quartile ratio \dqt{25}/\dqt{75} shows a similar measure of the
middle half of the spectrum.

For the $x$-axis in Fig.~7, one can simply use $m$
($\equiv\dqt{50}$) on a log scale to explore the wide range of
the hardness.  However, a simple log scale compresses the phase space
for relatively hard spectra ($m>0.5$; note $0\le m\le1$). Our 
expression\footnote{The inverse of hyperbolic tangent;  $\sim$ log($m$)
when $m \rightarrow$ 0 , and $\sim -\log(1-m)$ when $m \rightarrow$ 1.},
log($m/(1-m)$), shows both the soft and hard phase space equally well.

In Fig.~7, a flat spectrum lies at $x$=0 and $y$=1 in the diagram
($0<y<3$). The spectrum changes from soft to hard as one goes from left
to right in the diagram, and it changes from concave-upwards to
concave-downwards, as one goes from bottom to top.  The examples in the
previous sections explore a soft part of this phase diagram, which is
modeled by a power law.  

In the case of a narrow energy range, where an emission or absorption
line is expected on a relatively flat continuum, one can use the QCCD to
explore a spectral line feature even with limited statistics that normally
forbids normal spectral fitting.  In this case, the median ($x$-axis) in
the QCCD can be a good measure of line shift, and the quartile ratio
($y$-axis) can be a good measure of line broadening.

The right panel in Fig.~7 shows the overlay of the grid patterns for
typical spectral models - power law, thermal Bremsstrahlung, and black
body emission.  For high count sources, one can use such a diagram to
find relevant models for the spectra before detailed analysis.

\begin{figure*} \begin{center}
\epsscale{1.0}
\plotone{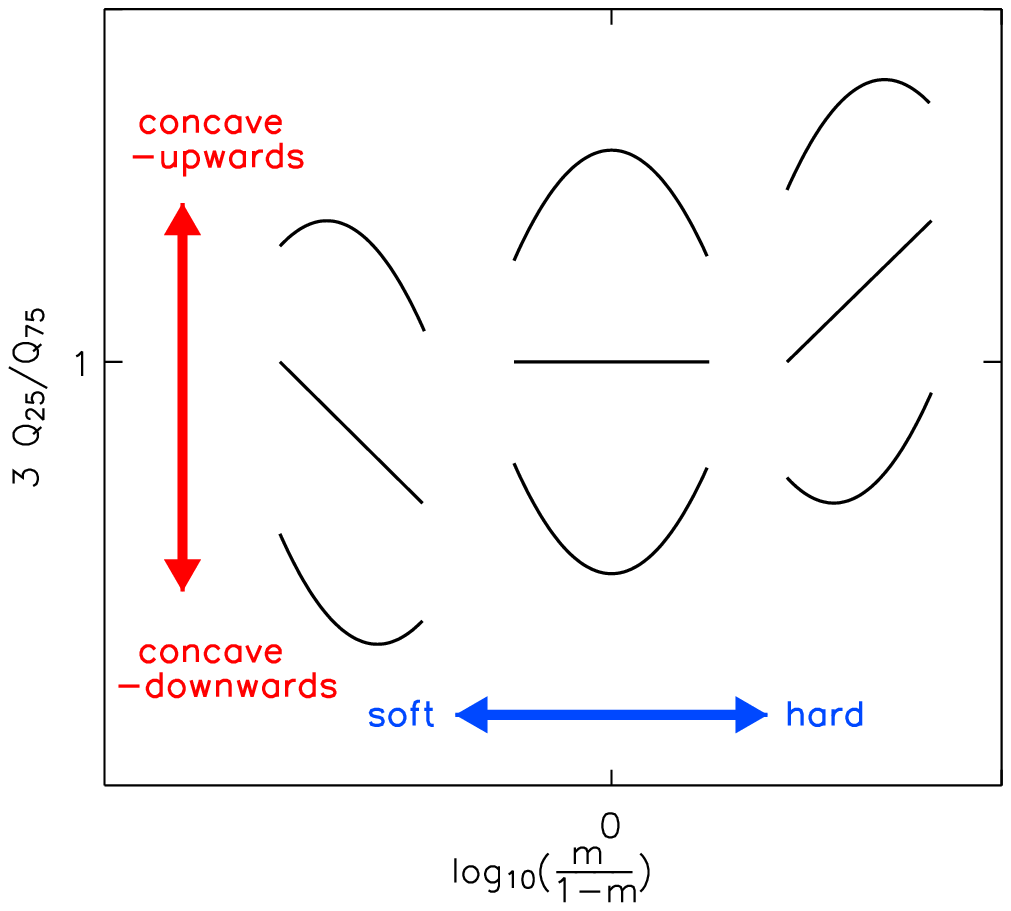}
\plotone{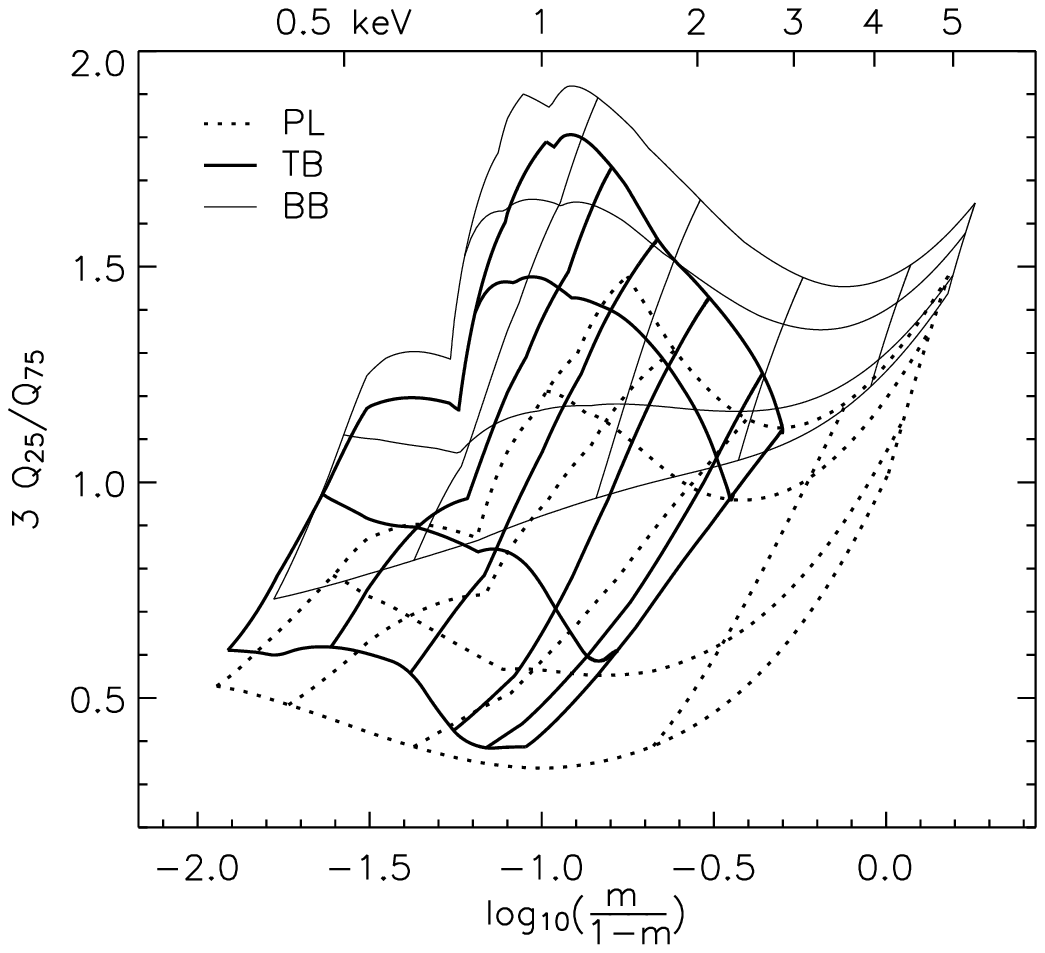}
\end{center} \caption{Overview of various spectral shapes 
(approximated as concave or straight lines) in the QCCD
using \dqt{50} and \dqt{25}/\dqt{75} (left), and the grid
patterns for power law (PL),
thermal Bremsstrahlung (TB), black body (BB) models in a 0.3-8.0 keV
range ideal detector  or incoming flux space
(right). For PL, the grid patterns are for \PLI= 4, 3, 2, 1 \& 0 along the
positive $x$-axis direction, for TB, $kT$ = 0.2, 0.5, 1, 2, 4, \& 10.0
keV, and for BB, $kT$ = 0.1, 0.2, 0.5, 1, 2, \& 4 keV.  For all the models, 
\nH= 0.1, 1, 5, \& 10 $\times 10^{21}$ cm\sS{-2} along the positive $y$
direction. } 
\label{fig:ov} \end{figure*}

Finally, we investigate how the detector energy resolution affects the
performance of the quantile-based diagram. The left panel in Fig.~8
shows the grid patterns of the power-law emission models for 
detectors with various energy resolutions but with the same detection
efficiency of the Chandra ACIS-S detector.  Starting with the energy
resolution of the Chandra ACIS-S detector in the previous examples
(Figs.~3, 4, 5 and 6; FWHM $\Delta E$ $\sim$ 150 eV at 1.5 keV and
$\sim$ 200 eV at 4.5 keV), we have successively decreased the energy
resolution by multiplying the energy resolution at
each energy by a constant factor.  Each pattern is labeled by the energy resolution ($\Delta
E/E$) at 1.5 keV.  As the energy resolution decreases, the grid pattern
shrinks.  The pattern will shrink down to a point ($x,y$)=(0,1) in the
diagram for a detector with no energy resolution.

The right panel in Fig.~8 shows the 68\% of the simulation results for a
detector with $\Delta E/E$ = 100\% at 1.5 keV. The size of the 68\%
distribution in this example is more or less similar to that in the case
of the regular Chandra ACIS-S detector with $\Delta E/E$ = 10\% (the
top-right panel in the Fig.~3).  The similarity is due to the fact that,
unlike the grid pattern, the dispersion of the simulation results (or
the error size of a data model point) for each model is mainly due to
statistical fluctuations for low count sources.  In summary, as the
energy resolution decreases, the relative distance between various
models in the diagram decreases, but the error size of the data remains
roughly the same, and thus the spectral sensitivity of the diagram
decreases.

Note that the overall size of the grid patterns in the left panel of
Fig.~8 is more or less similar when $\Delta E/E \lesssim 20$\% (at 1.5
keV). We expect that for a detector with energy resolution $\Delta E$,
the overall size is dependent on a quantity $\Delta
E/(E\Ss{up}-E\Ss{lo})$ since quantiles are determined over the full
energy range. Therefore, the quantile-based diagram is quite robust
against finite energy resolutions of typical X-ray detectors.

\begin{figure*} \begin{center} 
\epsscale{1.0}
\plotone{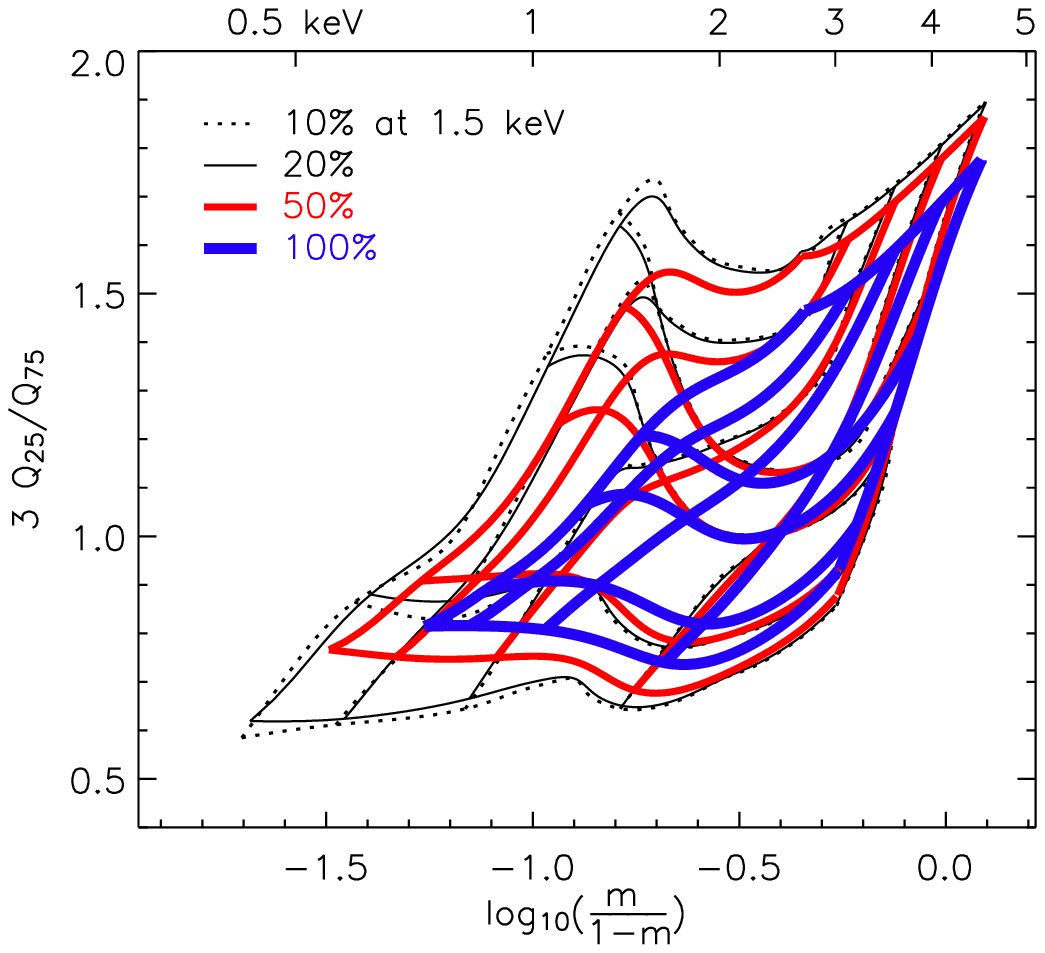}
\plotone{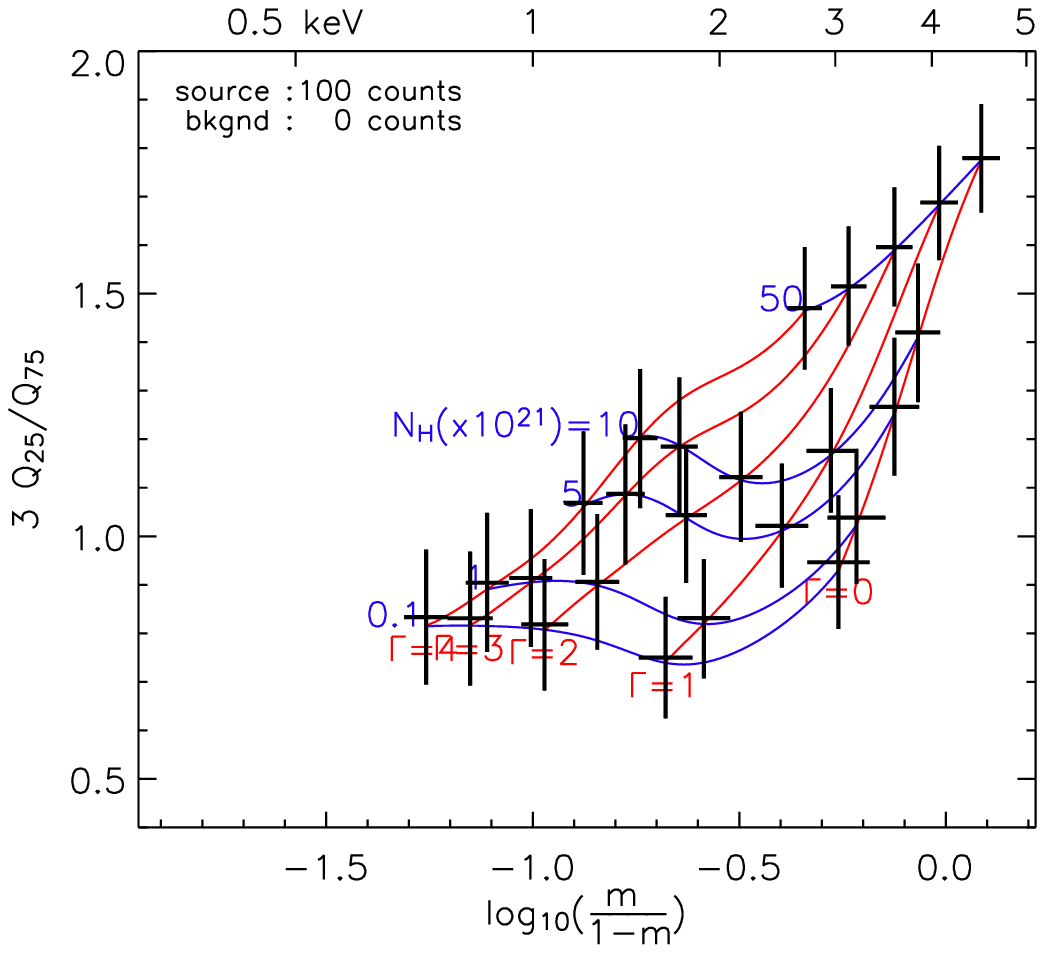} \end{center} \caption{ The grid patterns for various
energy resolutions with the same Chandra ACIS-S efficiency (left) and 
the 68\% distribution of the simulation results for 100 count sources
for a detector with $\Delta E/ E$ = 100\% ($\Delta E = $FWHM) at 1.5
keV (right).  Each grid pattern in the left panel is generated for
a different energy resolution, labeled by the energy resolution ($\Delta
E/ E$) at 1.5 keV.  As the resolution decreases, the pattern shrinks,
and if there is no energy resolution in the detector, the pattern
shrinks to a point at ($x,y$) = (0,1).  
The right figure shows the effects of limited photon statistics dominate
those of energy resolution (compare with Fig.~3, with $\Delta E/E = $10\%
at 1.5 keV) in determining the 68\% error bar sizes.
See the electronic ApJ for color version of the figure which distiguished
the model grids more easily.
} \label{fig:grid} \end{figure*}

\begin{figure*} \begin{center} 
\epsscale{1.0}
\plotone{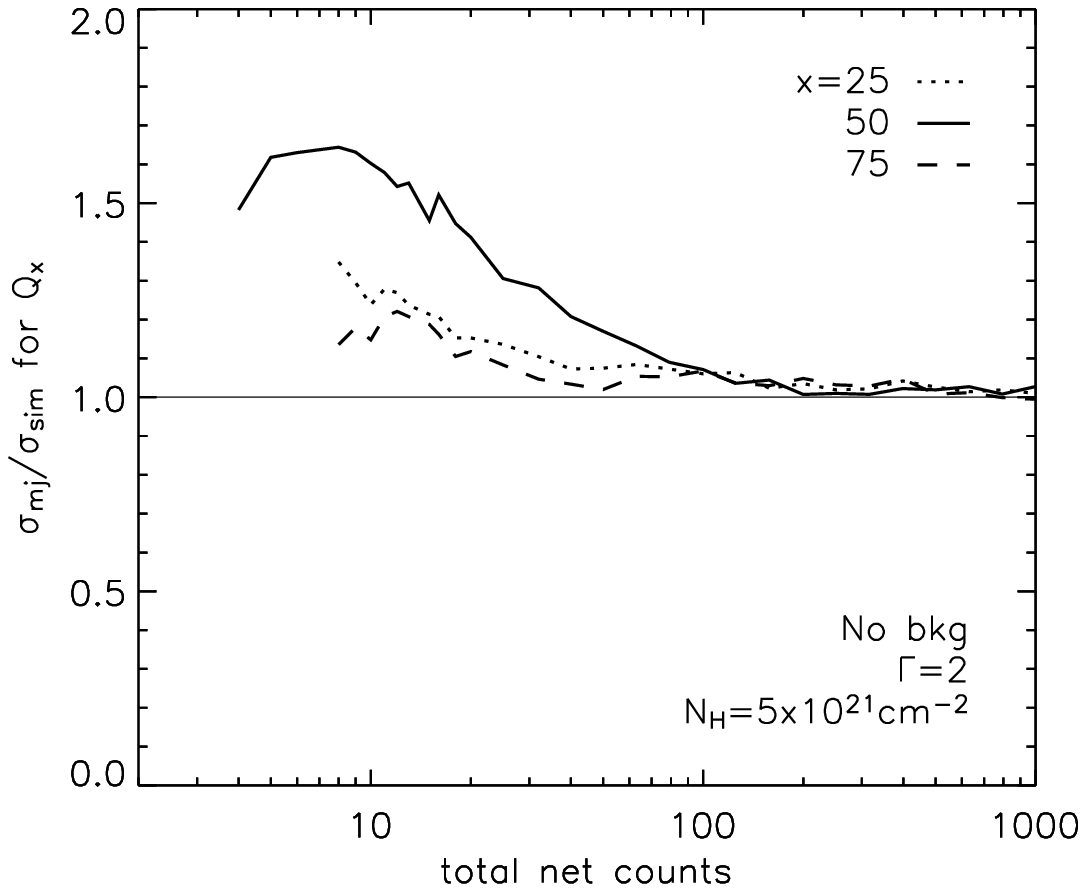}
\plotone{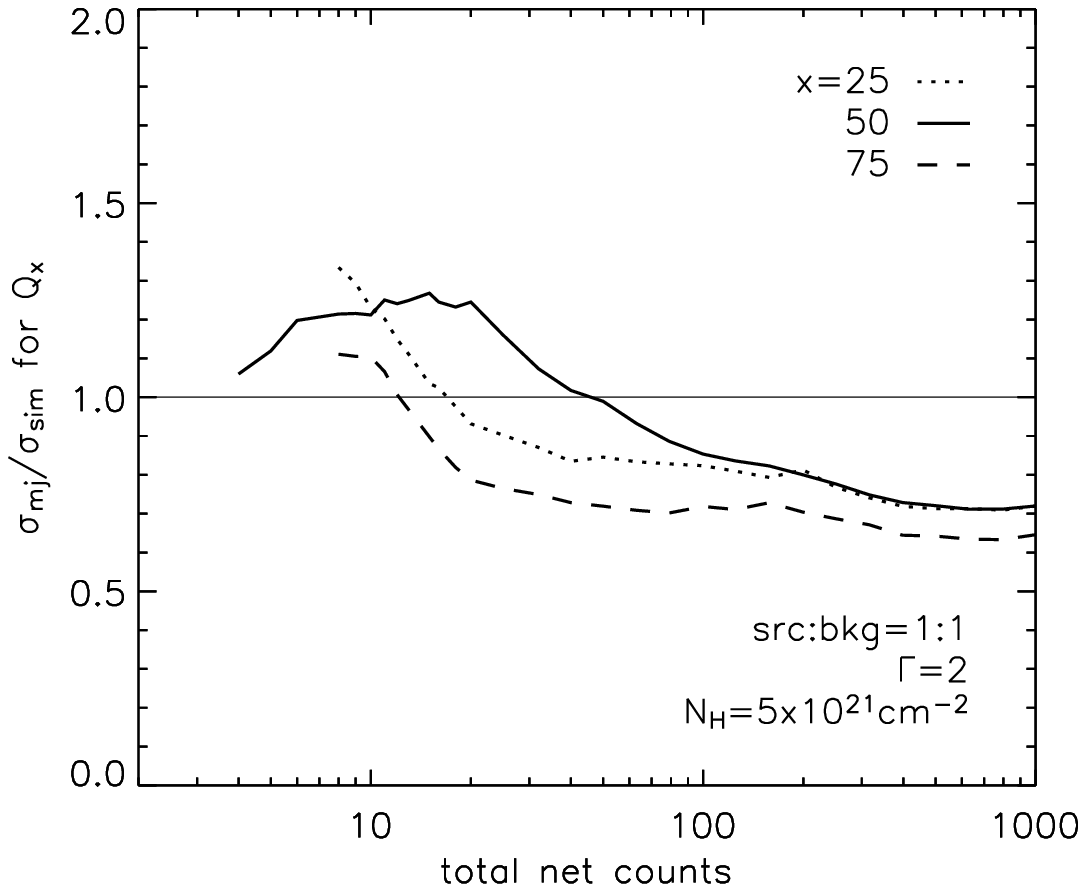}
\end{center} \caption{Accuracy of the error estimates by
the Maritz-Jarrett method: the ratio of the error estimate by the
Maritz-Jarrett method ($\sigma$\Ss{mj}) to that from simulations
($\sigma$\Ss{sim}, the 68\% distribution) for the cases with no background (left) and for the
cases with high background (right). See the ``background subtraction" section
in the appendix for the background subtraction routine.}
\label{fig:error} \end{figure*}

\section{Acknowledgement}
This work is supported by NASA grant AR4-5003A
for our on-going Chandra Multiwavelength plane (ChaMPlane)
survey of the galactic plane.

\appendix

Many routines have been developed to estimate quantiles and the related
statistics \citep{Wilcox97}. We use a simple interpolation technique
based on order statistics to estimate quantiles.  For real data, we also
need to have a reliable estimate for the errors of quantile values,
for which we employ the technique by \citet{Maritz78}.  For simplicity,
we assume that we measure the energy value of each photon.

\section{Quantile estimation using order statistics}

In the literature, many quantile estimation algorithms often
assume that the lowest value (energy) of the given distribution is the
(equivalent) lower bound of the distribution and the highest value
the upper bound.  In X-ray astronomy, the lower ($E$\Ss{lo}) and
upper bound ($E$\Ss{up}) of the energy range is usually set by the
instrument or user selection, where these bounds may or may not be the
lowest and highest energies of the detected photons. We can explicitly
impose this boundary condition by assigning 0\% and 100\% quantiles to
$E$\Ss{lo} and $E$\Ss{up} respectively.  

We sort the detected photons in ascending order
of energy, and we assign \qt{x} to the
energy $E_i$ of the $i$-th photon as 
\begin{eqnarray*} 
	E_i &=& E_{x\%}, \text{\ where \ }
	\frac{x}{100} = \frac{2 i-1}{2 N}, 
\end{eqnarray*}
and $N$ is the total number of net counts.  Using
$E$\Ss{lo}=$E_{0\%}$ and $E$\Ss{up}=$E_{100\%}$, one can interpolate any
quantiles from the above relation of $x$ and $E_i$.
With the definitions of $E$\Ss{lo}=$E_{0\%}$ and
$E$\Ss{up}=$E_{100\%}$, the above interpolation is very robust. One
can even calculate quantiles without any
detected photons (although not meaningful), the result of which is
identical to the case of a flat spectrum. 
In the case of only one detected photon at $E$, the
above relation reduces to $E$=$E_{50\%}$. Therefore, the distribution
of the median from one count sources with the same spectra is the
source spectrum itself.

The above interpolation for quantile \qt{x} essentially uses only two
energy values, $E_i$ and $E_{i+1}$, where $(2i-1)/2N < x/100 <
(2i+1)/2N$.  In order to take advantage of other energy values, one can
use more sophisticated techniques like that of \citet{Harrell82}.  In many
cases, the Harrell-Davis method estimates quantiles with smaller
uncertainties than the simple order statistics technique, but because of
smoothing effects in the Harrell-Davis method, there are cases that the
simple order statistics performs better, such as distributions
containing discontinuous breaks \citep{Wilcox97}.  In real data, the
finite detector resolution tends to smooth out any discontinuity
in the spectra.  Our simulation shows that about 10-15\% better
performance is achieved by the Harrell-Davis method compared to the simple order
statistics for the cases of 50 source photons and 25 background photons.
The results in the paper are generated by the order statistics
technique.

\section{Quantile error estimation using the Maritz-Jarrett method}

Once quantile values are acquired, one can always rely on simulations to
estimate their uncertainty (or error) using a suspected spectral model
with parameters derived from the QCCD. If no single model stands out as
a primary candidate, one can derive a final uncertainty of the quantiles
by combining the results of multiple simulations from a number of
models.   However, error estimation by simulations can be slow,
cumbersome, and even redundant, considering that the quantile errors are
more sensitive to the number of counts than the choice of spectral
shapes (cf. Figs.~3 and 4).  A quick and rough error estimate is often
sufficient, and so we introduce a simple quantile error estimate
technique -- the Maritz-Jarrett method.  The results (error bars and
shades in Fig.~1 to 6) in the main text indicate the interval that
encloses the central 68\% of the simulation results, and were not driven by
the Maritz-Jarrett method.

The Maritz-Jarrett method uses a technique similar to the
Harrell-Davis method of quantile estimation. Both methods rely on
L-estimators (linear sums of order statistics) using Beta functions.  
We sort the photons in the ascending order of energy from
$i=1$ to $N$.  
Then, we apply the Maritz-Jarrett method with small modifications\footnote{
Originally $M=[N x +0.5]$, where $[z]$ is the integer part of $z$.} as
follows. For \qt{x}, we set
\begin{eqnarray*} 
	M &=& N x +0.5 \\
	a &=& M-1 \\
	b &=& N-M.
\end{eqnarray*} We then define
$W_i$ using the incomplete beta function $B$,
\begin{eqnarray*} 
	W_i &=& B(a,b,y_{i+1}) - B(a,b,y_i) \\
	B(a,b,y) &=& \frac{\Gamma(a+b)}{\Gamma(a)\Gamma(b)}\int_0^y t^{a-1} (1-t)^{b-1} dt 
\end{eqnarray*} where $y_i = i/N$ and $\Gamma(n+1) = n!$ (the gamma
function).
Using $W_i$ and $E_i$, we calculate the L-estimators $C_k$,
\begin{eqnarray*} 
	C_k &=& \sum^{N}_{i=1} W_i E_i^k, \text{\ and then} \\
	\sigma(E_{x\%}) &=& \sqrt{C_2-C^2_1}. 
\end{eqnarray*}
The error estimate by the Maritz-Jarrett method requires at least 3
counts for medians, 5 counts for terciles, and 6 counts for quartiles
($a, b \ge1$) .

Fig.~9 shows the accuracy of the error estimates by the Maritz-Jarrett
method for \PLI = 2 and $\nH = 5\times10^{21}$ cm\sS{-2}.  In the case
of no background, the Maritz-Jarrett method is accurate when the total
net counts are greater than $\sim$ 30. It tends to overestimate the
errors when the total net counts are  below $\sim$ 30.  In the case of
high background, it tends to underestimate the errors overall since the
adopted background subtraction procedure (see below) does not inherit
background statistics for the error estimates.  We find that multiplying
by an empirical factor $\sqrt{1+N\Ss{bkg}/N\Ss{src}}$ can compensate,
approximately, for the underestimation ($N\Ss{src}$: total source
counts, $N\Ss{bkg}$: total background counts).

For the error of the ratio \dqt{25}/\dqt{75}, since these two quartiles
are not independent variables, the simple quadratic combination from two
quartile errors overestimates the error of the ratio ($\sim$ 20 -- 30\%
for the examples in Figs.~3 and 4).  One can find more sophisticated
techniques such as Bayesian statistics to estimate quantile errors in
the literature. For example, \citet{Babu99} and the reference therein
discuss the limitation of bootstrap estimation and show other techniques
such as the half-sample method.

\section{Background Subtraction}

For background subtraction, we calculate quantiles at a set of finely
stepped fractions separately for photons in the source region and the
background region.  Then, by a simple linear interpolation, we establish the 
integrated counts $IC(>$$E)$ (number of photons with energy greater than $E$) 
as a function of energy for both regions.  Now at each energy, one can
subtract the integrated counts of the background region from that of the
source region with a proper ratio factor of the area of the two regions.
Because of statistical fluctuations, the subtracted integrated net
counts may not be monotonically increasing from $E\Ss{lo}$ to
$E\Ss{up}$. Therefore we force a monotonic behavior by setting $IC(>$$E)
\ge IC(>$$E')$ if $IC(>$$E)< IC(>$$E') $ and $E>E'$.  Such a requirement
can underestimate the quantiles overall, so we repeat the above using
$IC(<$$E)$ and force a monotonic decrease. Then,  quantiles for the
net distribution are given by \begin{eqnarray*} x &=& \frac{N +
	 \mbox{$IC(>$$E)$} -\mbox{$IC(<$$E)$}}{2 N} \\ \qt{x}& =& E
\end{eqnarray*} where $N$ is the total net counts.  If there is no
statistical fluctuation, $IC(>$$E) = N - IC(<$$E)$.  For the error
estimation, we need to know the energy of each source photon, which will
be lost from background subtraction. So we generate a set of $N$ energy
values for source photons matching the above quantile relation and then
apply the above error estimation technique using the Maritz-Jarrett
method.  \end{document}